\documentstyle[12pt,a4]{article}
\begin{document}
\title{\sc Liouville Vortex And $\varphi^{4}$ Kink Solutions Of The 
Seiberg--Witten Equations}
\author{Serdar Nergiz and Cihan Sa\c{c}l\i o\~{g}lu\thanks{Permanent 
address: Physics Department, Bo\~{g}azi\c{c}i University, 
80815 Bebek--\.{I}stanbul, Turkey} }
\date{Physics Department, Bo\~{g}azi\c{c}i University \\
80815 Bebek--\.{I}stanbul, Turkey \\  and \\ Physics Department, TUBITAK  
\\  Marmara Research Center \\ 
Research Institute for Basic Sciences \\ 41470 Gebze, Turkey}
\maketitle
\vspace*{1cm} 
\begin{abstract}
The Seiberg--Witten equations, when dimensionally reduced to $\bf R^{2}\mit$, 
naturally yield the Liouville equation, whose solutions are parametrized by an 
arbitrary analytic function $g(z)$. The magnetic flux $\Phi$ is the integral of a 
singular Kaehler form involving $g(z)$; for an appropriate choice of $g(z)$ , $N$ 
coaxial or separated vortex configurations with $\Phi=\frac{2\pi N}{e}$ are 
obtained when the integral is regularized. The regularized connection in the 
$\bf R^{1}\mit$ case coincides with the kink solution of $\varphi^{4}$ theory.
\end{abstract}
\vspace*{5 cm}
\pagebreak
\baselineskip=30pt

The Seiberg--Witten equations \cite{SW} do not admit nonsingular solutions  
unless the curvature of the four dimensional base manifold $M$ happens  
to be negative over some regions of $M$. In particular,if $M=\bf R^{4}\mit$,  
the Weitzenbock formula implies that the modulus squared of the spinor field  
$\psi$ must either vanish everywhere, or exhibit singularities instead of local  
maxima \cite{KM}. There is also a global restriction on flat-space  
Seiberg--Witten solutions: Integrating the Weitzenbock formula, Witten  
\cite{SW} showed that all nontrivial flat solutions, including dimensionally 
reduced ones based on $\bf R^{3}$, $\bf R^{2}$ or $\bf R^{1}$, are all 
necessarily non-$L^{2}$.

Such singular, non-$L^{2}$ solutions, while probably not useful for Donaldson 
theory, may nevertheless be of physical interest. For example, Freund recently 
recognized that a singular  $U(1)$ magnetic monopole field and an 
accompanying spinor, found earlier by G\"{u}rsey in a different setting, solve 
the $\bf R^{3}$--reduced Seiberg--Witten equations \cite{F}. The monopole 
being the characteristic topological object in $\bf R^{3}\mit$, one may inquire 
whether there are $\bf R^{2}\mit$ and $\bf R^{1}\mit$ Seiberg--Witten solutions 
corresponding to vortices and kinks, respectively. The chief purpose of the 
present note is to show that such solutions indeed exist. A novel aspect of 
the $\bf R^{n}\mit$ $(n \leq 2)$ case is that the three coupled Seiberg--Witten 
equations (we should properly count $F=dA$ as one of the three) can be 
reduced to a single nonlinear one. This happens to be the Liouville equation 
\cite{LI}, which has recently been related to $N=2$ supersymmetric 
Seiberg--Witten theory in another context \cite{GO}.

We follow the conventions of Akbulut \cite{AK} in the choice of the Dirac 
$\gamma$--matrices
\begin{eqnarray}
\gamma^{1} &=& \left( \begin{array}{cc} \bf 0 \mit & \bf I \mit \\
                                             -\bf I \mit & \bf 0 \mit \end{array}\right)~~,~~
\gamma^{2} = \left( \begin{array}{cc} \bf 0 \mit & i\bf \sigma_{3}\mit \\
                   i\bf \sigma_{3}\mit & \bf 0 \mit \end{array}\right)~~, \nonumber \\
\gamma^{3} &=& \left( \begin{array}{cc} \bf 0 \mit & -i\bf \sigma_{2}\mit \\
                                 -i\bf \sigma_{2}\mit & \bf 0 \mit \end{array}\right)~~,~~
\gamma^{4} = \left( \begin{array}{cc} \bf 0 \mit & -i\bf \sigma_{1}\mit \\
                                    -i\bf \sigma_{1}\mit & \bf 0 \mit \end{array}\right)~~,
\label{1} 
\end{eqnarray}
and the self--dual $\Sigma_{ij}$
\begin{eqnarray}
\Sigma_{12}
& = &
\frac{1}{4}\{ [\gamma^{1} , \gamma^{2}]+[\gamma^{3} , \gamma^{4}] \}
 = 
\left( \begin{array}{cc} i\bf \sigma_{3}\mit & \bf 0 \mit \\
                                             \bf 0 \mit & \bf 0 \mit \end{array}\right)~~,
\nonumber \\
\Sigma_{13}
& = &
\frac{1}{4}\{ [\gamma^{1} , \gamma^{3}]+[\gamma^{4} , \gamma^{2}] \}
 = 
\left( \begin{array}{cc} -i\bf \sigma_{2}\mit & \bf 0 \mit \\
                             \bf 0 \mit & \bf 0 \mit \end{array}\right)~~, \label{2} \\
\Sigma_{23}
& = &
\frac{1}{4}\{ [\gamma^{1} , \gamma^{4}]+[\gamma^{2} , \gamma^{3}] \}
 = 
\left( \begin{array}{cc} -i\bf \sigma_{1}\mit & \bf 0 \mit \\
                                             \bf 0 \mit & \bf 0 \mit \end{array}\right)~~.
\nonumber
\end{eqnarray} 
Taking a spinor $\psi^{T}= (a\:,\:b\:,\:0\:,\:0)$, a connection $iA_{\mu}$ and its 
curvature $iF_{\mu\nu}=i(\partial_{\mu}A_{\nu}-\partial_{\nu}A_{\mu})$,
where $\mu,\nu=1,2,3,4~$, the first of the Seiberg--Witten pair is nothing but
the Dirac equation
\begin{equation}\label{3} 
\gamma^{\mu} (\partial_{\mu}+iA_{\mu}) \psi = 0~~.
\end{equation}
In the notation of \cite{AK}, the second Seiberg--Witten equation becomes
\begin{equation}\label{4} 
\rho(iF_{A}^{+})=\sigma(\psi)~~,
\end{equation}
where
\begin{equation}\label{5} 
\sigma(\psi) \equiv \left( \begin{array}{cc}
 \frac{(|a|^{2}-|b|^{2})}{2} & a\overline{b}  \\
                b\overline{a}& \frac{(|b|^{2}-|a|^{2})}{2}  
\end{array}\right)
\end{equation}
and
\begin{equation}\label{6} 
\rho(iF_{A}^{+})=\frac{i}{8}(F_{\mu\nu}+\tilde{F}_{\mu\nu})\cdot
\Sigma_{\mu\nu}= \frac{i}{4}F_{\mu\nu}\Sigma_{\mu\nu}~~.
\end{equation}
The vortex solutions in $\bf R^{2} \mit$ follow from the {\it Ansatz}
\begin{equation}\label{7} 
A_{\mu}=(A_{1}\:,\:A_{2}\:,\:0\:,\:0)~~,
\end{equation}
\begin{equation}\label{8} 
\psi^{T}= (a\:,\:b\:,\:0\:,\:0)~~,
\end{equation}
where all quantities are assumed to depend only on $x_{1}$ and $x_{2}$.
Putting (\ref{7}) and (\ref{8}) in (\ref{4}), one finds two possibilities: either 
$(a\neq0\:,\:b=0)$ or $(a=0\:,\:b\neq0)$. Choosing the first, (\ref{4}) reduces to
\begin{equation}\label{9} 
-F_{12}=-B_{3}=|a|^{2}
\end{equation}
while (\ref{3}) yields
\begin{equation}\label{10} 
(-\partial_{1}+i\partial_{2})a =(iA_{1}+A_{2})a~~.
\end{equation}
We now set
\begin{equation}\label{11} 
a =\alpha \exp (\omega_{x}+i\omega_{y})~~,
\end{equation}
where $\alpha$ is a constant with the dimensions of inverse length as 
required by (\ref{4}). This unusual dimension for the spinor field of course 
comes from the vacuum expectation value of the Higgs field in the twisted 
supersymmetric Yang--Mills theory underlying the Seiberg--Witten 
approach \cite{WSY}. Dividing both sides of (\ref{10}) by $a$, applying 
$(-\partial_{1}-i\partial_{2})$ and separating real and imaginary parts, we find 
\begin{equation}\label{12} 
(\partial_{1}^{2}+\partial_{2}^{2})\omega_{x}=-B_{3}=\alpha^{2} \exp (2\omega_{x})
\end{equation}
and
\begin{equation}\label{13} 
(\partial_{1}^{2}+\partial_{2}^{2})\omega_{y}=
-(\partial_{1}A_{1}+\partial_{2}A_{2})~~.
\end{equation}
In (\ref{12}), we have also used (\ref{9}). It is convenient to introduce the 
dimensionless coordinates $x=\alpha x_{1}$ , $y=\alpha x_{2}$ with 
$\alpha > 0$ and to define $z(\overline{z})=x+(-)iy$. The equation (\ref{12}) 
then becomes
\begin{equation}\label{14} 
4\partial_{z}\partial_{\overline{z}}\,\omega_{x}=\exp (2\omega_{x})~~.
\end{equation}
This is of course the well--known Liouville equation. Using (\ref{10}) and 
(\ref{11}) we obtain
\begin{equation}\label{15} 
A_{1}= \alpha (\partial_{y}\omega_{x}-\partial_{x}\omega_{y})
\end{equation}
and
\begin{equation}\label{16} 
A_{2}= -\alpha (\partial_{x}\omega_{x}+\partial_{y}\omega_{y})~~,
\end{equation}
which show that (\ref{13}) is automatically satisfied. (\ref{14}) has the solution 
\begin{equation}\label{17} 
\omega_{x}=\frac{1}{2}\ln\frac{4(dg/dz)(d\overline{g}/d\overline{z})}
{(1-g\overline{g})^{2}}~~,
\end{equation}
due to Liouville \cite{LI}. At this point, $g(z)$ is an arbitrary analytic function. 
Comparing (\ref{17}) with (\ref{11}), we see that
\begin{equation}\label{18} 
|a| =2\alpha\frac{\left| dg/dz \right|}{(1-g\overline{g})}~~,
\end{equation}
which naturally suggests
\begin{equation}\label{19} 
\omega_{y}=\pm\arg\frac{dg}{dz}=\mp\arg\frac{d\overline{g}}{d\overline{z}}~~.
\end{equation}
Note that this also makes $\omega_{y}$ harmonic and enforces 
$\vec{\nabla}\cdot\vec{A}=0$ via (\ref{13}). We finally take
\begin{equation}\label{20} 
a =2\alpha\frac{dg/dz}{(1-g\overline{g})}
\end{equation}
leading to
\begin{equation}\label{21} 
B_{3} =-\frac{4\alpha^{2}\left| dg/dz \right|^{2}}{(1-g\overline{g})^{2}}~~.
\end{equation}
The $U(1)$ curvature is thus seen to be the Kaehler $2$--form
\begin{eqnarray}
F
& = &
\frac{1}{e}F_{12}\:dx_{1}\wedge dx_{2}=\frac{i}{2e\alpha^{2}}
F_{12}\:dz\wedge d\overline{z}
\nonumber \\
& = &
\frac{-2i}{e}\frac{dg\wedge d\overline{g}}{(1-g\overline{g})^{2}}~~, \label{22} 
\end{eqnarray}
where we have brought out the coupling constant $e$ which was hidden in
$A_{\mu}$ all along. We can also combine (\ref{15}) and (\ref{16}) into the 
$1$--form
\begin{equation}\label{23} 
A =\frac{i}{e}\left( 
\frac{\overline{g}\:dg}{(1-g\overline{g})} - \frac{g\:d\overline{g}}{(1-g\overline{g})}
\right)~~.
\end{equation}
A number of remarks are in order:

{\bf (i)} The singularity dictated by the 
Weitzenbock formula manifests itself in (\ref{20})--(\ref{23}). The solutions are
singular in the z--plane along a curve defined by $g\overline{g}= 1$. One can  
easily trace the minus in $(1-g\overline{g})$ to the relative plus sign between 
the two sides of the Liouville equation (\ref{14}); introducing a relative minus 
sign in (\ref{14}) changes all the $(1-g\overline{g})$ factors to  
$(1+g\overline{g})$ without affecting anything else. {\bf (ii)} Remarkably, 
(\ref{14}) and (\ref{22}) also arise in a twice-dimensionally reduced Ansatz 
leading to vortexlike solutions of the self--dual Yang--Mills equations 
\cite{SAC}. Since the  $\bf R^{4}\mit$ self--dual Yang--Mills system is 
conjectured to generate all integrable systems through various dimensional 
reductions, the appearance of (\ref{14}) in both the SDYM and the  
Seiberg--Witten contexts may be regarded as an additional clue for similar  
integrability properties of the latter. Furthermore, putting $e^{w_{x}}=u$ in the 
Liouville equation (\ref{14}) and performing a further dimensional reduction by 
demanding $u=u(|z|=r)$ results in the differential equation for the 3rd 
Painlev\'{e} transcendent with $\gamma=1$, $\alpha=\beta=\delta=0$, in 
the notation of Ince \cite{IN}. Thus the Seiberg--Witten equations also exhibit 
a Painlev\'{e} property, considered an indication of integrability \cite{PN}. 
{\bf (iii)} In the SDYM case, passing from the $(++++)$ $\bf R^{4}\mit$ to the 
twistor--based $(++--)$ $\bf R^{2,2}\mit$ supplies the change in the relative 
sign in (\ref{14}) converting the singular $(1-g\overline{g})^{-1}$ factor into 
$(1+g\overline{g})^{-1}$; the same obviously holds in our problem as well. 
{\bf (iv)} The curvature form (\ref{22}) remains unchanged under 
$g\rightarrow 1/g$, just as it would under a gauge transformation. Indeed, 
this inversion of $g$ precisely gives rise to the $U(1)$ gauge transformations 
\begin{equation}\label{24} 
a \rightarrow a'  =\frac{\overline{g}}{g}a
\end{equation}
and 
\begin{equation}\label{25} 
eA \rightarrow eA'  =eA+id\ln\frac{g}{\overline{g}}
\end{equation}
on the spinor field and the connection, respectively.

We now wish to restrict the choice of $g(z)$ by physical considerations. In 
Freund's case, $\int F$ gives the quantized magnetic charge; it would be 
natural to expect that $\int F\equiv\Phi$ is a quantized magnetic flux in  
$\bf R^{2}\mit$ for an appropriate $g(z)$. This requires that we somehow 
``regularize'' the singular integrand (\ref{22}); happily, different approaches 
to making sense out of  $\int F$ gives the same result, as we shall see. Let us 
start by considering the $\bf R^{2,2}\mit$ version of  (\ref{22}), which becomes 
the area of the Riemann sphere stereographically expressed onto the 
g--plane:
\begin{equation}\label{26} 
\Phi= \int F = \frac{2i}{e}\int\frac{dg\wedge d\overline{g}}{(1+g\overline{g})^{2}}~.
\end{equation}
Note the overall sign change due to the change in the RHS of (\ref{14}). 

Trying $g=z^{\nu}$ for an axisymmetric solution centered at the origin, we find 
\begin{equation}\label{27} 
\Phi= \frac{4\pi\nu}{e}\int_{0}^{\infty}\frac{(2\nu r^{2\nu -1})}{(1+r^{2\nu})^{2}}dr=
\frac{4\pi\nu}{e}\int_{1}^{\infty}\frac{dw}{w^{2}}=\frac{4\pi\nu}{e}~~,
\end{equation}
where we have put $w=1+r^{2\nu}$. We note that the gauge transformation 
(\ref{24}) results in
\begin{equation}\label{28} 
a' = \overline{z}^{\nu}z^{-\nu}a= e^{-2i\nu\theta}a~~.
\end{equation}
The singlevaluedness of $a'$ at $\theta = 2\pi$ allows the $\nu$ values 
\begin{equation}\label{29} 
\nu=\frac{1}{2},\;1,\;\frac{3}{2},\ldots ~~.
\end{equation}
We need not consider $\nu\rightarrow -\nu$ as this only amounts to the 
gauge transformation $g\rightarrow 1/g$ mentioned earlier. Of the values in 
(\ref{29}), it is the $\nu=\frac{1}{2}$ that corresponds to $2\pi /e$, i.e., the 
Nielsen--Olesen \cite{NO} unit of flux. Thus $g=z^{1/2}$ represents the basic 
single--vertex solution, while $g=z^{n/2}$ corresponds to a single vortex with 
$n$ units of flux. It is now easy to verify that
\begin{equation}\label{30} 
g(z)= \prod_{k=1}^{n}(z-a_{k})^{1/2}
\end{equation}
describes n vortices centered at the locations $a_{k}=(a_{kx}+ia_{ky})$. To do  
this, we first switch to the compactified version of  (\ref{23}), which becomes
\begin{equation}\label{31} 
A =-\frac{i}{e}\left( 
\frac{\overline{g}\:dg}{(1+g\overline{g})} - \frac{g\:d\overline{g}}{(1+g\overline{g})}
\right)~~.
\end{equation}
Next we use the $g(z)$ of (\ref{30}) in 
\begin{equation}\label{32} 
\Phi=\int_{\bf R^{2}\mit}F = \int_{\partial\bf R^{2}\mit}A~~,
\end{equation}
where $\partial\bf R^{2}$ is a clockwise circle whose radius goes to infinity. 
Since $|g|\rightarrow\left| z^{n/2} \right|\gg 1$ on ${\partial\bf R^{2}\mit}$, we 
obtain
\begin{equation}\label{33} 
\Phi=-\frac{i}{2e}\oint_{\partial \bf R^{2}\mit}\left\{ n \frac{dz}{z}-
 n\frac{d\overline{z}}{\overline{z}}\right\}=\frac{2n\pi}{e}~~.
\end{equation}
The similarity of expression (\ref{30}) to Weierstrassian functions suggests we 
might consider a doubly--periodic solution on a two dimensional lattice, with 
one vortex per unit lattice cell. Let us take $\omega_{1}$ and $\omega_{2}$  
as the two basic lattice vectors, subject to the usual restriction 
$Im (\omega_{2} / \omega_{1})\neq 0$. For a pair of integers 
$(n_{1},\:n_{2})$, $\omega=n_{1}\omega_{1}+n_{2}\omega_{2}$ is a point in the 
lattice. We can now choose for $g(z)$ the square root of the Weierstrassian 
quasi--periodic function, i.e., 
\begin{equation}\label{34} 
g(z)= \sigma^{1/2}(z)=z^{1/2}\prod_{\omega\neq 0}(1-\frac{z}{\omega})^{1/2}
\exp \left( \frac{z^{2}}{4\omega^{2}}+\frac{z}{2\omega} \right)~~.
\end{equation}
The exponential factor is needed to ensure the convergence of the product. 

Another method for defining the integral of the singular expression (\ref{22}) is 
as follows. If we attempt to calculate $\Phi$ starting from (\ref{22}), we obtain 
\begin{equation}\label{35} 
\Phi=-\frac{4\pi\nu}{e}\int_{0}^{\infty}\frac{(2\nu r^{2\nu-1})}{(r^{2\nu}-1)^{2}}dr=
-\frac{4\pi\nu}{e}\int_{-1}^{\infty}\frac{dw}{w^{2}} \; \; \; (w=r^{2\nu}-1)
\end{equation}
instead of (\ref{27}). Adopting Speer's analytic regularization \cite{SP}, we 
define 
\begin{equation}\label{36} 
I(-1,\:\infty)=\int_{-1}^{\infty}\frac{dw}{w^{2}}=\left[ \int_{-1}^{\infty}w^{\lambda}dw
\right]_{\lambda=-2} = -1~~.
\end{equation}

This is equivalent to writing $I(-1,\:\infty)=I(-\infty,\:\infty)-I(-\infty,\:-1)$ and  
throwing away the infinite ``constant'' $I(-\infty,\:\infty)$. Thus we get the same 
answer as in the compactified $\bf R^{2,2}\mit$ formulation.

Now let us return to equation (\ref{8}) and ask what happens if we take 
$a=0$, $b\neq 0$. It is easy to check that one still ends up with the Liouville 
equation (\ref{14}); the changes consist of $B_{3}\rightarrow -B_{3}$ and 
\begin{equation}\label{37} 
b =2\alpha\frac{(d\overline{g}/d\overline{z})}{(1-g\overline{g})}~~.
\end{equation}
Thus while it is not possible to change the direction of the magnetic field by 
$g\rightarrow 1/g$, anti--vortices can be obtained by 
$(a(g),\:0)\rightarrow(0,\:b(\overline{g}))$, $B_{3}\rightarrow -B_{3}$.

Finally, let us briefly examine the $n=1$ case. A possible Ansatz is 
\begin{equation}\label{38} 
A_{\mu}=(0\:,\:0\:,\:0\:,\:A_{4}(x_{1}))
\end{equation}
and
\begin{equation}\label{39} 
\psi^{T}= (a(x_{1})\:,\:b(x_{1})\:,\:0\:,\:0)
\end{equation}
The Seiberg--Witten equation (\ref{6}) demands either 
$\psi^{T}= (a\:,\:a\:,\:0\:,\:0)$ or $\psi^{T}= (a\:,\:-a\:,\:0\:,\:0)$. Taking 
$a =\alpha \exp (\omega_{x}+i\omega_{y})$ as before, these two cases yield
\begin{equation}\label{40} 
\partial_{1}(\omega_{x}+i\omega_{y})= \pm A_{4}
\end{equation}
respectively. Thus in order for $A_{4}$ be real, $\omega_{y}$ can at most be 
a constant, which we may take to be zero. Then, using (\ref{3}) we obtain 
\begin{equation}\label{41} 
\partial_{1}^{2}\omega_{x}= \alpha^{2}\exp (2\omega_{x})
\end{equation}
Calling $ \alpha x_{1} = x$ again, (\ref{41}) is seen to be a $y$--independent 
version of the Liouville equation (\ref{12}). While (\ref{41}) may be integrated 
directly by elementary methods, it is simpler to read off the solution from 
(\ref{17}) by picking a $g(z)$ such that the variable $y$ disappears in 
$\omega_{x}$. This happens only for $g(z) = \exp \kappa (z+x_{0})$, $\kappa$ 
and $x_{0}$ being constant real numbers (an imaginary constant added to 
$x_{0}$ cancels out along with the $iy$ in (\ref{17})). We may as well set 
$x_{0} = 0$, which gives
\begin{equation}\label{42} 
\omega_{x}=\frac{1}{2}\ln\frac{4\kappa^{2}e^{2\kappa x}}{(1-e^{2\kappa x})^{2}}~~.
\end{equation}
This results in
\begin{equation}\label{43} 
|a| =\frac{\alpha\kappa}{\left| \sinh\kappa x \right|}~~,
\end{equation}
\begin{equation}\label{44} 
A_{4} =\pm \alpha\kappa\coth\kappa x ~~,
\end{equation}
and 
\begin{equation}\label{45} 
E_{1} =\mp \frac{\alpha^{2}\kappa^{2}}{ \sinh^{2}\kappa x }~~.
\end{equation}
The expected singularity appears at $x = 0$ in (\ref{42})--(\ref{45}). In contrast,  
the non-singular version obtained by $x_{1}\rightarrow ix_{1}$ has 
$(1+e^{2\kappa x})^{-2}$ in (\ref{42}); in addition, $\cosh\kappa x $ and 
$\sinh\kappa x $ in (\ref{43})--(\ref{45}) are now switched. Thus $A_{4}$ 
changes into $\pm \alpha\kappa\tanh\kappa x $, which is the well--known 
kink (antikink) solution of $\varphi^{4}$ theory.

In conclusion, we see that the dimensionally reduced Seiberg--Witten 
equations in $\bf R^{n}\mit$ $(n= 1\:,\:2\:,\:3)$ yield singular version of 
topological solitons characteristic of each $n$, the $n= 3$ case being 
represented by Freund's monopole solution. The accompanying spinors are 
of course the new feature associated with these familiar solitons. The $n= 2$ 
case indicates connections between integrable systems and the 
Seiberg--Wittten equations. Finally, it should be interesting to look for 
solutions of the Seiberg--Witten equations reduced to a two--dimensional 
manifold admitting negative local values for the scalar curvature and see how 
this affects the singularities.

C.S. gratefully acknowledges useful conversations with A.H. Bilge, S. Finashin 
and S. Akbulut, whom he also thanks for providing ref.\cite{AK}


\begin{thebibliography}{99}

\bibitem{SW} E. Witten,  
Math. Research Letters {\bf 1} (1994) 769.
\bibitem{KM} P. Kronheimer and T. Mrowka,  
Math. Research Letters {\bf 1} (1994) 797.
\bibitem{F} P.G.O. Freund, 
J. Math. Phys. {\bf 36} (1995) 2673 ; 
F. G\"{u}rsey, in {\it Gauge Theories and Modern Field Theory}, edited by 
R. Arnowitt and P. Nath (MIT, Cambridge, 1976), p. 369
\bibitem{LI} J. Liouville, 
J. Math. Appl. {\bf 18} (1853) 71. 
\bibitem{GO} A. Gorsky, I. Krichever, A. Marshakov, A. Mironov and 
A. Morozov, ``Integrability and Seiberg--Witten Exact Solution'', 
HEP--TH--9505035. 
\bibitem{AK} S. Akbulut,
Journal reine angew. Math {\bf 447} (1994) 83 ; 
``Lectures on Seiberg--Witten Invariants'' (unpublished).
\bibitem{WSY} E. Witten,
J. Math. Phys. {\bf 35} (1994) 5101.
\bibitem{SAC} C. Sa\c{c}l\i o\~{g}lu , 
J. Math. Phys. {\bf 25} (1984) 3214.
\bibitem{IN} E.L. Ince, 
{\it Ordinary Differential Equations}, (Dover, New York, 1956) p.345
\bibitem{PN} M.J. Ablowitz, A. Ramani and H. Segur,
J. Math. Phys. {\bf 21} (1980) 715.
\bibitem{NO} H. Nielsen and P. Olesen,
Nucl. Phys. {\bf B 61} (1973) 45.
\bibitem{SP} E. Speer,
Ann. Math. Stud. {\bf 62} (1969).

\end{thebibliography}
\end{document}